\documentclass[aps,prc,twocolumn,showpacs]{revtex4}
\usepackage{ulem}
\usepackage{color}
\usepackage{graphicx}
\usepackage{epsfig}
\usepackage{bm}
\usepackage{amssymb}

\def\nn {\nonumber}

\newcommand{\be}{\begin{equation}}
\newcommand{\ee}{\end{equation}}
\newcommand{\bea}{\begin{eqnarray}}
\newcommand{\eea}{\end{eqnarray}}

\newcommand{\om}{\omega}
\newcommand{\vk}{\vec k}
\newcommand{\vq}{\vec q} 
\newcommand{\mn}{\mu\nu}

\begin{document}
\title{Transport coefficients in finite volume Polyakov--Nambu--Jona-Lasinio model}
\author{Kinkar Saha$^{1}$, Sabyasachi Ghosh$^{2}$, Sudipa Upadhaya$^{3,4}$, Soumitra Maity$^{3,4}$} 
%
\affiliation{$^1$Department of Physics, University of Calcutta, 92, 
A. P. C. Road, Kolkata - 700009, India}
\affiliation{$^2$Indian Institute of Technology Bhilai, GEC Campus, Sejbahar, Raipur-492015, 
Chhattisgarh, India}
\affiliation{$^3$Center for Astroparticle Physics $\&$ Space Science,
Block-EN, Sector-V, Salt Lake, Kolkata 700091, West Bengal, India}
\affiliation{$^4$ Department of Physics, Bose Institute, 93/1, A. P. C. Road,
Kolkata 700009, West Bengal, India}

\begin{abstract}
Finite size consideration of matter significantly affects
transport coefficients like shear viscosity, bulk viscosity, electrical conductivity,
which we have investigated here in the framework of the 
Polyakov--Nambu--Jona-Lasinio model.
Owing to the basic quantum mechanics, a non-zero lower momentum cut-off
is implemented in momentum integrations, used in the expressions of
constituent quark masses and transport coefficients. 
When the system size decreases, the values of these transport coefficients
are enhanced in low temperature range. At high temperature domain, shear
viscosity and electrical conductivity become independent of system sizes.
Whereas, bulk viscosity, which is
associated with scale violating quantities of the system, faces some non-trivial
size dependence in this regime. In the phenomenological direction, our microscopic
estimations can also be linked with the macroscopic outcome, based on dissipative 
hydrodynamical simulation.
\end{abstract}

\maketitle

\section{Introduction}
\label{sec:intro}
%
%
The experiment of relativistic heavy ion collider (RHIC) has created a nearly 
perfect fluid~\cite{Romatschke:2007mq,Luzum:2008cw,Roy:2012jb}, whose 
shear viscosity to entropy density ($\eta/s$) ratio is so small that it
almost reaches the lower bound ($\eta/s=1/4\pi$)~\cite{KSS}. 
In the high temperature domain however, the theoretical calculations
using perturbative methods surprisingly do not lead to such small value 
of $\eta/s$. There it behaves as weakly interacting gas, having relatively large
value (10-20 times 
larger than lower bound)~\cite{Arnold}.
To resolve this discrepancy
between experimental and theoretical values,
different alternative calculations, based on effective
QCD models~\cite{Purnendu,Redlich_NPA,Marty,G_CAPSS,Weise2,Kinkar,G_IFT,Deb,Tawfik}
and hadronic 
models~\cite{ Itakura,Dobado,Nicola,Weise1,SSS,Ghosh_piN,Gorenstein,HM,Kadam:2015xsa,Hostler}  
have been studied in recent times. Some estimations are also done from the direction
of transport simulations~\cite{Bass,Muronga,Plumari,Pal} and Lattice 
QCD calculations~\cite{Meyer_eta,LQCD_eta2}.
Other transport coefficients
like bulk viscosity ($\zeta$)~\cite{{Paech1},{Gavin},{Arnold_bulk},{Prakash},{Tuchin},
{Tuchin2},{Nicola},{Marty},{De-Fu},{Redlich_NPA},{Redlich_PRC},{G_IFT},{Deb},{Tawfik},
{Purnendu},{Vinod},{Santosh},{Sarkar},{HM},{Kadam:2015xsa},{Kadam:2015fza},
{Sarwar:2015irq},{Hostler},{Nicola_PRL},{SG_NISER},{Meyer_zeta},{Dobado_zeta1},{Dobado_zeta2},{Saha:2015lla}}
, electrical conductivity ($\sigma$) ~\cite{LQCD_Ding,LQCD_Arts_2007,LQCD_Buividovich,
LQCD_Burnier,LQCD_Gupta,LQCD_Barndt,LQCD_Amato,Cassing,Puglisi,Greif,
{Marty},{PKS},{Finazzo},{Lee},{Nicola_PRD},{Greif2},{G_IFT2},{Ghosh:2016yvt}}
of this QCD medium have also become matter of contemporary interest.
From these earlier research, we get 
a gross summary about the temperature dependence of these transport coefficients.
$\eta(T)$ and $\sigma(T)$ decrease and increase in temperature domains of
hadron and quark phases respectively,
while $\zeta(T)$ follows an opposite trend. So, near 
transition temperature, one can expect a maximum 
in $\zeta(T)$~\cite{Redlich_NPA,De-Fu,Purnendu,Saha:2015lla} and
minimum in $\eta(T)$~\cite{Purnendu,Redlich_NPA,G_CAPSS,Weise2,Kinkar,Deb} 
and $\sigma(T)$~\cite{Cassing}. 

These analysis were carried out for infinite size systems. 
Effects of finiteness in system volumes have not however been considered which we 
are studying in this work.
We know that the lower bound of
shear viscosity to entropy density ratio ($\eta/s$) is basically lowest possible quantum fluctuation
of fluid, which can never vanish, even in the infinite coupling limit~\cite{KSS}. 
However, in this infinite coupling limit, we can think about a classical fluid, whose $\eta/s\rightarrow 0$.
On the other hand, $\eta/s$ of RHIC matter is surprisingly close to its quantum lower bound, 
which indicates that the matter is very sensitive to the quantum fluctuations. Hence other 
possible quantum effect like finite size may be important to be considered.
%
There is a long list of Refs.~\cite{Ferdinand:1969zz,Fisher:1972zza, Luscher:1985dn, 
Elze:1986db, Gasser:1987ah, Spieles:1997ab,
Gopie:1998qn, Kiriyama:2002xy, Fischer:2005nf, Abreu:2006pt, Shao:2006gz, Yasui:2006qc,
Bazavov:2007ny, Palhares:2009tf, Luecker:2009bs, Braun:2010vd, Fraga:2011hi, Braun:2011iz,
Abreu:2011rj, Abreu:2011zzc, Ebert:2011tt, Khanna:2012zz, Bhattacharyya:2012rp,%
Bhattacharyya:2014uxa, Bhattacharyya:2015zka, Bhattacharyya:2015kda, Magdy:2015eda,
Redlich:2016vvb,RedlichV, Xu:2016skm,Bhattacharyya:2015pra}, where finite size effect on 
different physical quantities have been investigated.
From experimental point of view as well, the produced fireball might have a finite system
volume, depending on the size of the colliding nuclei, the center of mass energy
and centrality of the collision.
Significances of these issues raise immediate question
on its impact over the transport coefficients of a system.
In this manuscript, we intend to explore the same in a qualitative manner.
For this initial work, we do so by implementing a lower momentum cut-off
following the same line of studies as in~\cite{Bhattacharyya:2015kda}.
Our studies have been carried out
within the realm of Polyakov--Nambu--Jona-Lasinio (PNJL) model
incorporating upto six quark type of interactions.
This finite volume effect on transport coefficients is recently 
investigated in HRG model by Samanta et al.~\cite{Sub_HRG} and Sarkar et al.~\cite{Nachiketa}, 
which is valid
for hadronic temperature domain only. Here we have
explored this fact in PNJL model, which can well describe the thermodynamics of
QCD medium for entire temperature domain, which contain
quark and hadron both phases. This model also additionally
contains the finite volume effect of quark condensate, for which a major
change in transport coefficients is observed.

The article is organized as follows. Next section has
covered the finite system size picture of PNJL model and then a brief 
formalism part of transport coefficients.
Then our numerical outputs are analyzed in the result section
and at last, we summarize our studies.


\section{The model framework}
The framework that we shall be working
with is that of 2+1 flavor PNJL model~\cite{Meisinger,Fukushima1,Ratti1,
Megias1,Ghosh1,SMukherjee1,Ghosh2,Tsai1,Megias2}. This model entwines
two very basic features of QCD viz. chiral symmetry and
its spontaneous breaking and confinement physics. The quark
dynamics is incorporated in the NJL part through
multi-quark interaction terms. Here we shall consider upto six-quark
type of interactions. The gluon dynamics on the other hand, is taken
care of through a
background field representing Polyakov loop dynamics. 
There has been a considerable progress made in this direction
in order to understand properly the strongly interacting system under
this framework~\cite{Marty,Kinkar,Saha:2015lla,Abuki1,Abuki2,
Boomsma1,Bhat4,Sasaki1,Bhat5,Bhat7,KSaha1,Dutra1,Ghosh3,
Xin1,Islam1,Ghosh4,Islam2,Bhattacharyya:2012rp,Bhattacharyya:2014uxa,
SuKi,hybrid}.

The Polyakov loop potential~\cite{Saumen} is expressed as,
\begin{equation}
 \frac{\mathcal{U}'[\Phi,\bar{\Phi},T]}{T^4}=
 \frac{\mathcal{U}[\Phi,\bar{\Phi},T]}{T^4}-
 \kappa \ln \big( J[\Phi,\bar{\Phi}] \big)
 \label{Ploop-potential1}
\end{equation}
where, the second term on the right hand side is the Vander-monde 
term~\cite{Ghosh2} reflecting the effect of SU(3) Haar measure.
$\mathcal{U}[\Phi,\bar{\Phi},T]$ is the Landau-Ginzburg
type potential chosen to be of the form,
\begin{equation}
 \frac{\mathcal{U}[\Phi,\bar{\Phi},T]}{T^4}=
 -\frac{b_2(T)}{2}\bar{\Phi}\Phi-\frac{b_3}{6}
 \big( \Phi^3 +\bar{\Phi}^3 \big)+
 \frac{b_4}{4}\big( \Phi\bar{\Phi}\big)^2
 \label{Ploop-potential2}
\end{equation}
The coefficients $b_3$ and $b_4$ are kept constant, whereas
the temperature dependence is included in $b_2$ with a form,
\begin{equation}
 b_2(T)=a_0+a_1\bigg( \frac{T_0}{T}\bigg)+a_2\bigg( \frac{T_0}{T}\bigg)^2
 +a_2\bigg( \frac{T_0}{T}\bigg)^3
 \label{ploop-potential3}
\end{equation}
All the associated parameters are set~\cite{Saumen} through few physical
constraints and the rest by fitting with available results from
Lattice QCD. The set of parameters that we have chosen for present 
purpose can be found in~\cite{Bhattacharyya:2015kda}.

For the quark dynamics, we shall use similar framework of NJL model
except replacing with co-variant derivative in the kinetic part 
of the Lagrangian in
presence of Polyakov loop. Under 2+1-flavor consideration with upto 
six-quark type of interactions, the Lagrangian gets modified as
is given in~\cite{Saumen}. As a result of dynamical breaking of chiral
symmetry in the NJL model, the chiral condensate $\langle\bar{\Psi}\Psi\rangle$
acquires non-zero vacuum expectation values. The constituent mass
as a consequence is given by,
\begin{equation}
 M_f~=~m_f-g_S\sigma_f+g_D\sigma_{f+1}\sigma_{f+2}
 \label{NJL1}
\end{equation}
where $\sigma_f\equiv\langle \bar{\Psi}_f\Psi_f \rangle$ represents the 
chiral condensate. If $\sigma_f=\sigma_u$, then $\sigma_{f+1}=\sigma_d$
and $\sigma_{f+2}=\sigma_s$ and in cyclic order. 

Now in order to implement the effect of finite system sizes, one is ideally supposed 
to choose the proper boundary conditions : periodic for bosons and anti-periodic for
fermions. This in effect leads to a sum of infinite extent over discretized 
momentum values, $p_i=\frac{\pi n_i}{R}$, $R$ being the dimension of cubical volume.
$n_i$ are positive integers with i=x,y,z. This would then imply an infra-red cut-off
$p_{min}=\frac{\pi}{R}=\lambda (\rm say)$. Ideally the surface
and curvature effects should be taken care of as well. 
However, this being the very first case study in this direction, we are mostly 
interested in the qualitative 
changes of the transport coefficients under finite system size consideration.
To obtain that, we incorporate few simplifications.
The infinite sum over discrete momentum values
will be replaced by integration over continuum momentum variation, albeit with the 
infra-red cut-off. Alongside, we are not going to use any amendments in the mean-field
values due to finite system sizes. This in effect implies that the system volume, $V$ will be
regarded as a parameter just like temperature, T and chemical potential, $\mu$ on the 
same footing. Parametrization will be the same as for zero T, zero $\mu$ and infinite V.
Any variation therefore occurring due to any of these parameters will be reflected in
$\sigma_f$, $\Phi$ etc. and through them in meson spectra.

With these simplifications, the thermodynamic potential thereafter takes the form,
\begin{eqnarray}
\centering
&& \Omega=
  \mathcal{U'}(\Phi[A],\bar{\Phi}[A],T)
 +2g_S\sum_{f=u,d,s}\sigma_f^2-\frac{g_D}{2}\sigma_u\sigma_d\sigma_s
\nonumber \\
 &-& 6\sum_f\int_\lambda^\Lambda\frac{d^3p}
  {(2\pi)^3}E_{p_f}\Theta(\Lambda-|\vec{p}|)
       - 2\sum_f T\int_\lambda^\infty\frac{d^3p}{(2\pi)^3}
\nonumber \\
 &&\ln \left[1+3\left(\Phi+\bar{\Phi}
            e^{\frac{-(E_{p_f}-\mu_f)}{T}}\right)
            e^{\frac{-(E_{p_f}-\mu_f)}{T}}            
           +  e^{\frac{-3(E_{p_f}-\mu_f)}{T}} \right]
\nonumber \\
       &-& 2\sum_f T\int_\lambda^\infty\frac{d^3p}{(2\pi)^3}
       \nonumber \\
  &&\ln \left[1+3\left(\Phi+\bar{\Phi}
            e^{\frac{-(E_{p_f}+\mu_f)}{T}}\right)
            e^{\frac{-(E_{p_f}+\mu_f)}{T}} 
           + e^{\frac{-3(E_{p_f}+\mu_f)}{T}}\right]
\end{eqnarray}
where, each term bears its usual significance, which can be found in~\cite{Saumen}.

\section{transport coefficients}
\label{sec:formal}
%
Green-Kubo relation~\cite{Zubarev,Kubo} connects transport
coefficients like shear viscosity $\eta$, bulk viscosity $\zeta$ 
and electrical conductivity $\sigma$
to their respective thermal fluctuation or correlation
functions - $\langle\pi^{ij}(x)\pi_{ij}(0)\rangle_\beta$,
$\langle{\cal P}(x){\cal P}(0)\rangle_\beta$
and $\langle J^i(x)J_i(0)\rangle_\beta$, 
where $\langle ..\rangle_\beta$ stands for thermal average. 
The operators $\pi_{ij}$ and ${\cal P}$ 
can be obtained from energy-momentum
tensor $T^{\mn}$ 
as,
\bea
\pi^{ij}&\equiv&T^{ij}-g^{ij}T^k_k/3~,
\nn\\
{\cal P}&\equiv&-T^k_k/3 -c_s^2T^{00}
\eea 
where $c_s$ is speed of sound in the medium. 
The operator $J^i$ is the electromagnetic current of medium constituents.
The transport coefficients in momentum space can be written 
explicitly in spectral representations as,
\bea
\eta&=&\frac{1}{20}\lim_{q_0,\vq\rightarrow 0}
\frac{\int d^4xe^{iq\cdot x}\langle [\pi^{ij}(x),\pi_{ij}(0)]\rangle_\beta}{q_0}~,
\nn\\
\zeta&=&\frac{1}{2}\lim_{q_0,\vq\rightarrow 0}
\frac{\int d^4xe^{iq\cdot x}\langle [{\cal P}(x),{\cal P}(0)]\rangle_\beta}{q_0}~,
\nn\\
\sigma&=&\frac{1}{6}\lim_{q_0,\vq\rightarrow 0}
\frac{\int d^4xe^{iq\cdot x}\langle [J^{i}(x),J_{i}(0)]\rangle_\beta}{q_0}
\eea

Our aim of this work is to calculate these transport coefficients of quark matter 
under the framework of PNJL model and to notice their changes because of the finite size
consideration of medium. As we know that the mathematical expressions of transport coefficients,
calculated from relaxation time approximation (RTA) in kinetic theory approach
and the one-loop diagram in quasi-particle Kubo approach are exactly same,
let us start with the standard expressions of $\eta$~\cite{Nicola,Weise1,G_IJMPA}, 
$\zeta$~\cite{Nicola,G_IFT}
and $\sigma$~\cite{Nicola_PRD,Nicola,G_CU} given by, 
\bea
\eta&=&\frac{g}{15 T}\int \frac{d^3\vk}{(2\pi)^3}
\tau_{Q}\left(\frac{\vk^2}{\om_{Q}}\right)^2[f^{+}_{Q}(1-f^{+}_{Q})
\nn\\
&& + f^{-}_{Q}(1-f^{-}_{Q})]~;
\label{eta_G}
\eea
\bea
\zeta&=&\frac{g}{T}\int \frac{d^3\vk}{(2\pi)^3\om_{Q}^2}
\tau_{Q}\left\{\left(\frac{1}{3}-c_s^2\right)\vk^2 
 - c_s^2 m_Q^2
\right.\nn\\
&&\left. ~~ -c_s^2m_QT\frac{dM_Q}{dT}\right\}^2\left[ f^{+}_{Q}\Big(1-f^{+}_{Q}\Big ) 
\right.\nn\\
&&\left. ~~~~~~~~~~~~ + f^{-}_{Q}\Big (1-f^{-}_{Q}\Big )\right]~;
\label{zeta_Gmu0}
\eea
\bea
\sigma&=&\frac{6{\tilde e}_Q^2}{3 T}\int \frac{d^3\vk}{(2\pi)^3}
\tau_{Q}\left(\frac{\vk}{\om_{Q}}\right)^2[f^{+}_{Q}(1-f^{+}_{Q})
\nn\\
&&~~~~~~~~~~~~~~~~~~~~~ + f^{-}_{Q}(1-f^{-}_{Q})]~,
\label{sigma_G}
\eea
where $g=2\times 2\times 3$
is degeneracy factor of quark,
$f_Q^{\pm}$ are the modified Fermi-Dirac (FD) distribution functions of quarks and anti-quarks
respectively in presence of Polyakov loop. $\om_Q=\{\vk^2 +m_Q^2\}^{1/2}$ is the
single quasi-particle energy and
\be
{\tilde e}_Q^2=\left\{\left(+\frac{2}{3}\right)^2 + \left(-\frac{1}{3}\right)^2 \right\}e^2~.
\ee
%
\section{Numerical results and discussions}
\label{sec:num}
\begin{figure} 
\begin{center}
\includegraphics[width=0.5 \textwidth]{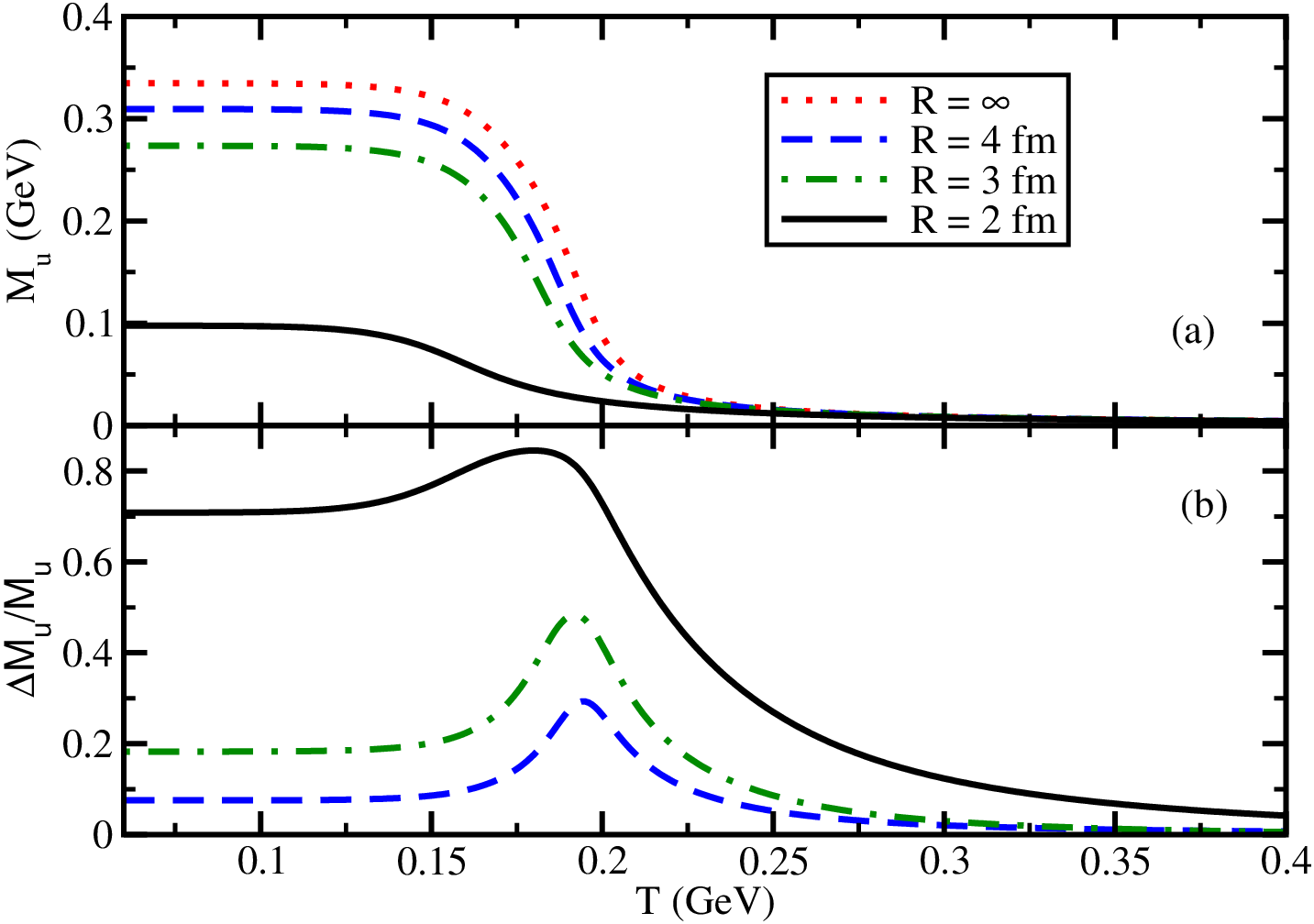}
\caption{(Color online) (a) $T$ dependence of $u$ (or $d$) quark mass for 
$R=\infty$ (red dotted line), $4$ fm (blue dashed line),
$3$ fm (green dash-dotted line) and $2$ fm (black solid line).
(b) Change of the $u$ (or $d$) quark mass for the same 
finite size parameters with respect to the quark mass for $R=\infty$.} 
\label{m_ud_T}
\end{center}
\end{figure}
\begin{figure} 
\begin{center}
\includegraphics[width=0.5 \textwidth]{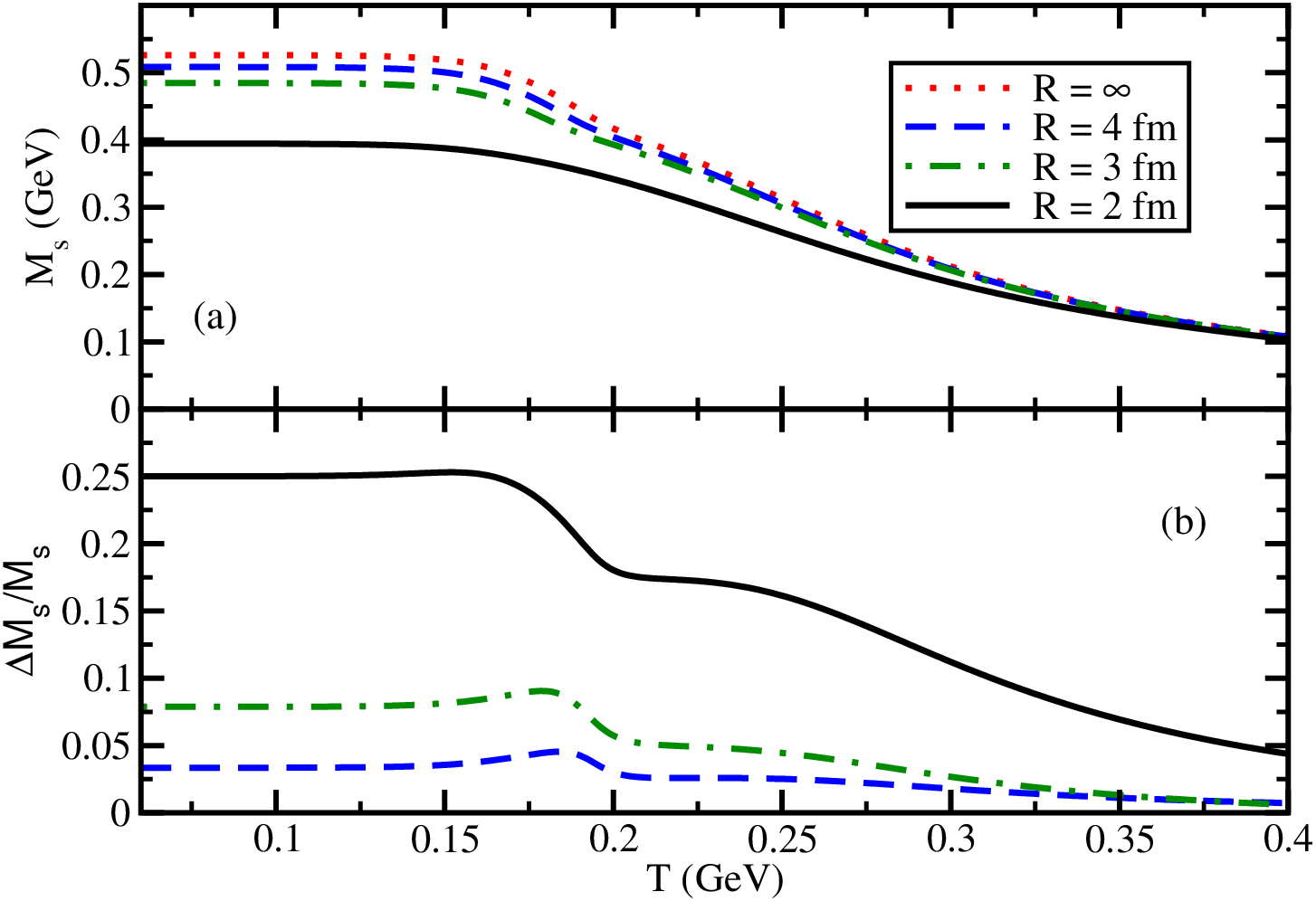}
\caption{(Color online) Same as Fig.~(\ref{m_ud_T}) for strange quark.} 
\label{m_s_T}
\end{center}
\end{figure}
%
%
Let us first take a glance at the expressions of transport coefficients, 
given in Eqs.~(\ref{eta_G}), (\ref{zeta_Gmu0}) and (\ref{sigma_G}). 
Replacing the $\vk=0$ by $\vk=\pi/R$ in the lower limit of the integrations,
we have adopted the effect of finite size of quark matter, having the dimension 
$R$. Along with this, there is another place in Eqs.~(\ref{eta_G}), 
(\ref{zeta_Gmu0}) and (\ref{sigma_G}), where the finite size effect 
 enters. This is the integrand part of transport coefficients, which
depends on the constituent quark mass and thus changes due to consideration
of finite size of quark matter. So there are two sources from where finite size
effect will modify our numerical estimations of transport coefficients. 

Before analyzing the numerical outputs, let us discuss about the limitations
of our present formalism. In finite temperature quantum field theory, we have to
introduce the imaginary time parameter, which can vary from $\tau=0$ to $\tau=-i\beta$.
This finite time restriction makes the energy component discretized via
Matsubara prescription (imaginary time formalism). Similarly, restriction of 
finite size or length ($L$) makes the momentum component discretized
i.e. the four momentum variables $(k_0, \vk)$ will be discretized
as $k_0\rightarrow \om_n=\frac{2\pi}{\beta}(n+1/2)$ and 
$\vk\rightarrow \vk_n=\frac{2\pi}{L}(n+1/2)$, where $n=0,\pm 1, \pm 2,...$
because of finite temperature $T=1/\beta$ and length $L$~\cite{Abreu:2011rj}.
Following analytic continuation technique, the discrete sum 
of energies can be transformed to its continuous integration.
For three momentum component, their discrete sum is roughly assumed as
continuous integration, starting from the lower momentum cutoff. 
This simplified picture of finite size effect by 
implementing lower momentum cut-off is justified in Refs.~\cite{Redlich:2016vvb,RedlichV}.
It is nicely demonstrated in Fig.~(1) of Ref.~\cite{Redlich:2016vvb}.
Along with the discretization of momentum, surface and curvature effects may
appear in the finite size picture~\cite{Shao:2006gz} but we don't consider
these effects in the present work for simplicity.

Let us now start our numerical discussion from the size dependency of constituent
quark mass, which is shown in Figs.~(\ref{m_ud_T}) and (\ref{m_s_T}).
As can be seen from Eq.~(\ref{NJL1}), the temperature dependent condensates determine the temperature
dependence of constituent quark masses, which are plotted by red dotted lines in 
Figs.~\ref{m_ud_T}(a) and \ref{m_s_T}(a) for u and s quarks. These are the results
of $M_u(T)$ and $M_s(T)$ when we have not considered any finite size effect.
We have marked this result by $R=\infty$. 
Now, introducing finite size consideration in gap equation (Eq.~(\ref{NJL1})), we get the curves
- blue dashed line, green dash-dotted line and black solid line for $R=4$ fm, $3$ fm,
$2$ fm respectively in Figs.~\ref{m_ud_T}(a) and \ref{m_s_T}(a). To zoom in the changes
of masses due to finite size effect, we have defined,
\be
\frac{\Delta M_{u,s}}{M_{u,s}}=\frac{M_{u,s}-M_{u,s}(R)}{M_{u,s}}~,
\label{del_M}
\ee
where  $M_{u,s}(R)$ are u, s quark masses for the medium with dimension $R$
and $M_{u,s}$ for $R=\infty$. 
These $\frac{\Delta M_{u,s}}{M_{u,s}}$ are plotted in Figs.~\ref{m_ud_T}(b) 
and \ref{m_s_T}(b) for $R=4$ fm, $3$ fm and $2$ fm.
We see that $\frac{\Delta M_{u,s}}{M_{u,s}}$ increases as $R$ decreases,
shows a mild peak before melting down at
high temperature. The peaks mainly appear because of
faster melting rates of M(R=$\infty$) compared to the rest.
%
%
The vanishing nature of $\frac{\Delta M_{u,s}}{M_{u,s}}$
at high $T$ limit is expected because of gradual restoration of chiral symmetry
in that regime for any system size. So, one may safely ignore the finite size effect of quark gluon plasma(QGP), produced
in heavy ion experiments, when it remains very hot, far above transition temperature, $T_c$. The finite size 
effect of QCD medium becomes important
near $T_c$ and its non-perturbative (hadronic) temperature domain. This general feature of 
finite size effect is observed for various quantities, 
discussed later.
\begin{figure} 
\begin{center}
\includegraphics[width=0.48 \textwidth]{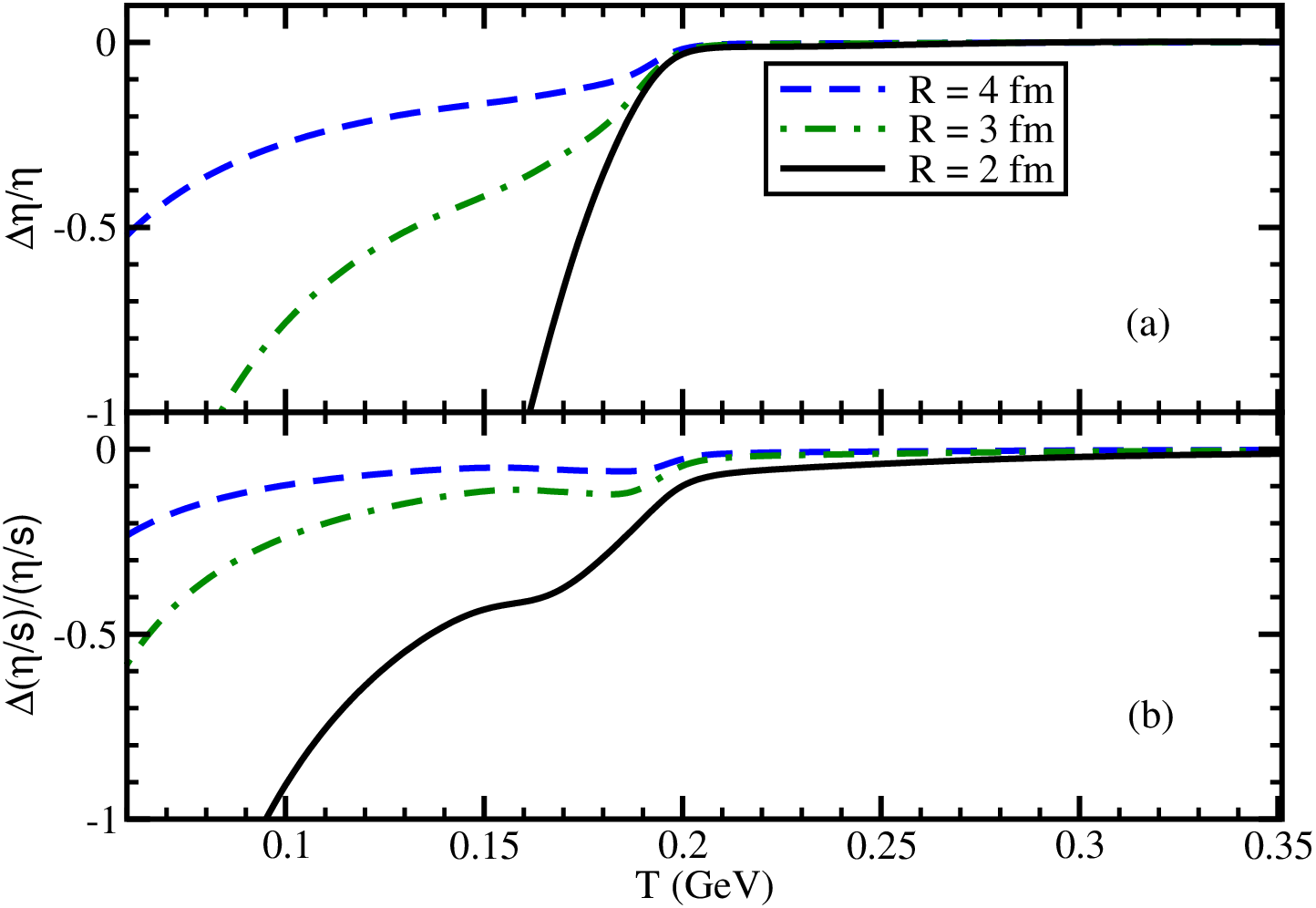}
\caption{(Color online) Difference between the results of finite and infinite $R$
for the quantities - shear viscosity $\eta$ (a) and shear viscosity to entropy
density ratio $\eta/s$ (b).} 
\label{eta_T}
\end{center}
\end{figure}
\begin{figure} 
\begin{center}
\includegraphics[width=0.5 \textwidth]{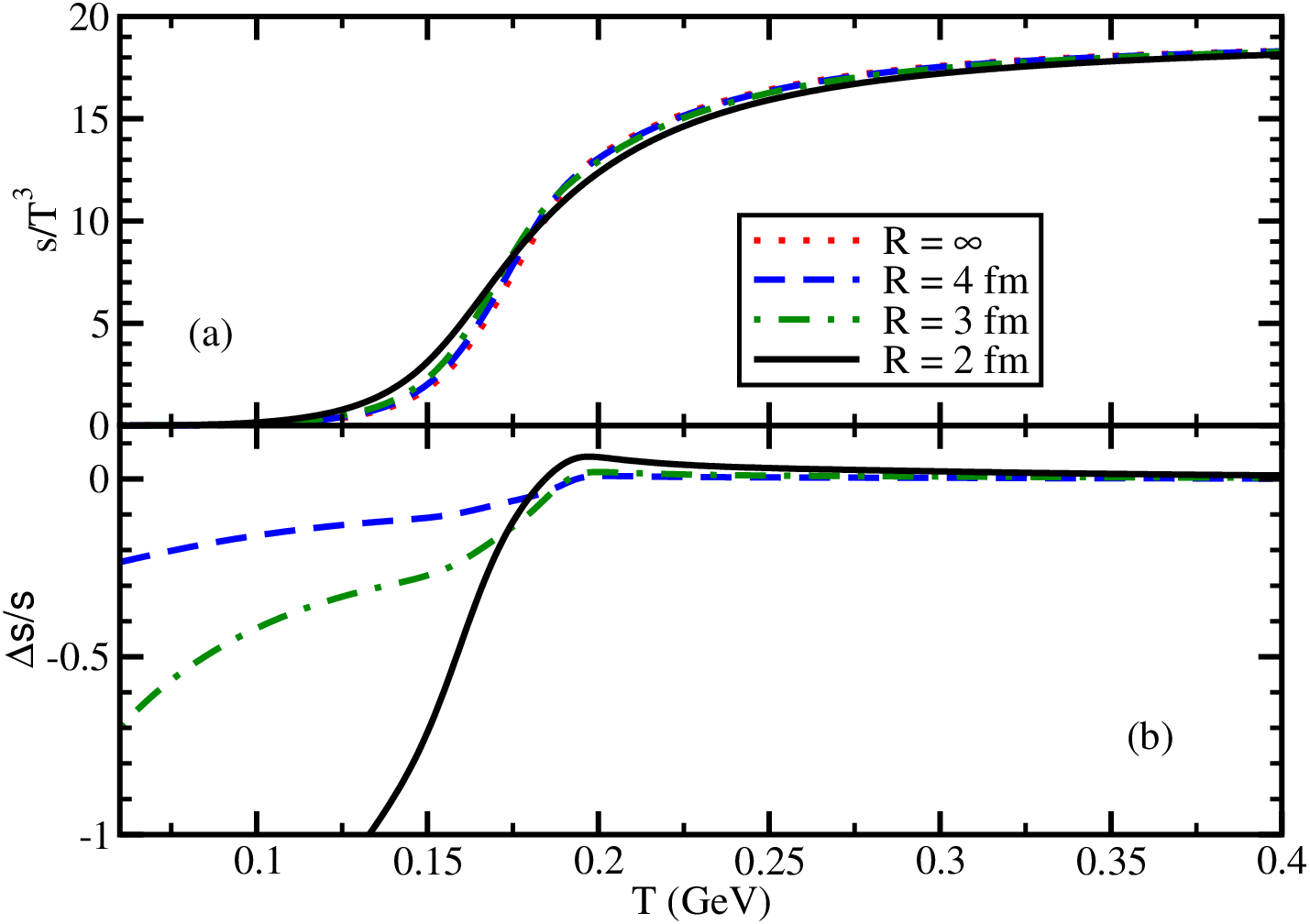}
\caption{(Color online) (a) $T$ dependence of entropy density, normalized by $T^3$ for different
values of $R$. (b) Difference between finite and infinite matter results for entropy
density.} 
\label{s_T}
\end{center}
\end{figure}

Now, using the expression of $M_{u,s}(T,R)$ and lower momentum cutoff $\vk=\pi/R$ in Eq.~(\ref{eta_G}),
we can generate $\eta(T)$ for different values of $R$.
Following same definition, given in Eq.~(\ref{del_M}), we have defined $\Delta \eta/\eta$,
which are plotted in Fig.~\ref{eta_T}(a). The negative values of $\Delta \eta/\eta$ below
$T=200$ MeV indicate that shear viscosity gets enhanced because of finite size effect. In one side,
$\eta$ should be decreased because of lower momentum cutoff in Eq.~(\ref{eta_G}), while
on other side, reduction of constituent quark mass for finite $R$ will act on the integrand
part of Eq.~(\ref{eta_G}) to enhance the values of $\eta$. Latter source dominates over the former
one, therefore, a net enhancement of $\eta(T<200$ MeV$)$ is observed in our results. 
Next, to discuss the finite size effect on $\eta/s$, shown in Fig.~\ref{eta_T}(b), let us
focus on entropy density $s$, obtained from the thermodynamical potential $\Omega$.
Figs.~\ref{s_T}(a) and (b) show the $T$ dependences of $s/T^3$ and $\Delta s/s$ for different
values of $R$. 
When this finite
size effect of $s$ at low $T$ enters into the quantity $\eta/s$, a less amount of enhancement of $\eta/s$
has been found with respect to the enhancement of $\eta$. For example, at $T=170$ MeV and $R=2$ fm in
Fig.~(\ref{eta_T}), we see that $70\%$ enhancement in $\eta$ shrinks to $40\%$ enhancement in $\eta/s$. 
Also, for vanishing chemical potential, we observe only cross-over transitions,
which is true for all $R$. So no discontinuity in $\eta$ or $\eta/s$ is
observed. However, with decreasing $R$, $T_c$ is supposed to decrease 
~\cite{Bhattacharyya:2015kda} and that
is visible in Fig.~\ref{eta_T}(b). The locations of slight bending there, grossly
representing the transition region, shift towards lower $T$ with decrease in $R$.
\begin{figure} 
\begin{center}
\includegraphics[width=0.5 \textwidth]{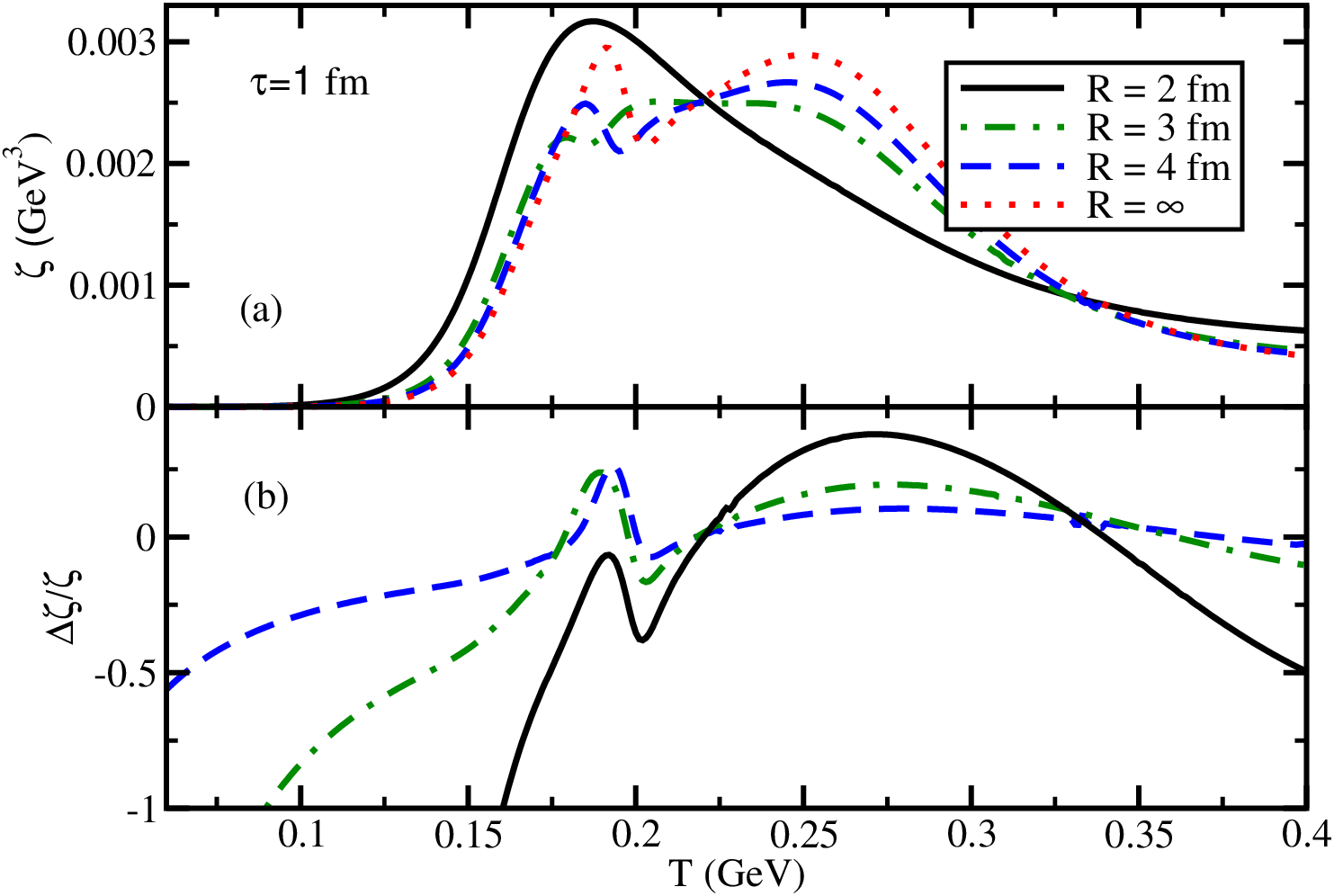}
\caption{(Color online) (a) $\zeta(T)$ for $R=\infty$ (red dotted line), $4$ fm (blue dashed line),
$3$ fm (green dash-dotted line) and $2$ fm (black solid line). (b) Difference between 
finite and infinite matter results for $\zeta$.} 
\label{zeta_T}
\end{center}
\end{figure}

Let us come to the next transport coefficient - bulk viscosity $\zeta$,
which is a very interesting quantity because of its relation with conformal
symmetry of the system.
Compared to Eq.~(\ref{eta_G}) for $\eta$, $\zeta$ in Eq.~(\ref{zeta_Gmu0})
contains additionally a conformal breaking term 
\be
\left\{\left(\frac{1}{3}-c_s^2\right)\vk^2  - c_s^2 \left(M_Q^2
+M_QT\frac{dM_Q}{dT}\right)\right\}^2~,
\label{conf}
\ee
which vanishes in the limits of $c_s^2\rightarrow 1/3$ and $M_{Q}\rightarrow 0$.
The QCD matter at high temperature can achieve these limits, where it behaves as 
a scale independent or conformally symmetric system. 
It can alternatively be realized
from the vanishing values of $\zeta$ for this QCD matter at high temperature.
Relating to this fact, our bulk viscosity estimation in PNJL model is trying to measure indirectly 
the breaking of this conformal symmetric nature of QCD matter in both quark and hadronic domain. 
In this context, the present investigation has tried to explore the
finite system size effect on this breaking of conformal symmetry
by studying the $R$ dependence of $\zeta$. For a constant value of relaxation time
($\tau=1/\Gamma=1$ fm), we have estimated $\zeta(T)$ for $R=\infty$ (red dotted line), 
$4$ fm (blue dashed line), $3$ fm (green dash-dotted line) and $2$ fm (black solid line), 
which are drawn in Fig.~\ref{zeta_T}(a). We see double peak-like structures in $\zeta$,
which are also observed in other earlier calculations~\cite{Dobado_zeta2,G_IFT}.
%
%
These double peak-like structures start diluting as we decrease system
sizes and for $R=2$ fm such nature disappears, indicating some non-trivial 
contributions from strange sectors. As we know, with decreasing system sizes,
the constituent masses acquire smaller values in the low temperature domain,
thereby 
tending towards restoring the chiral symmetry over the entire temperature window.

\begin{figure} 
\begin{center}
\includegraphics[width=0.5 \textwidth]{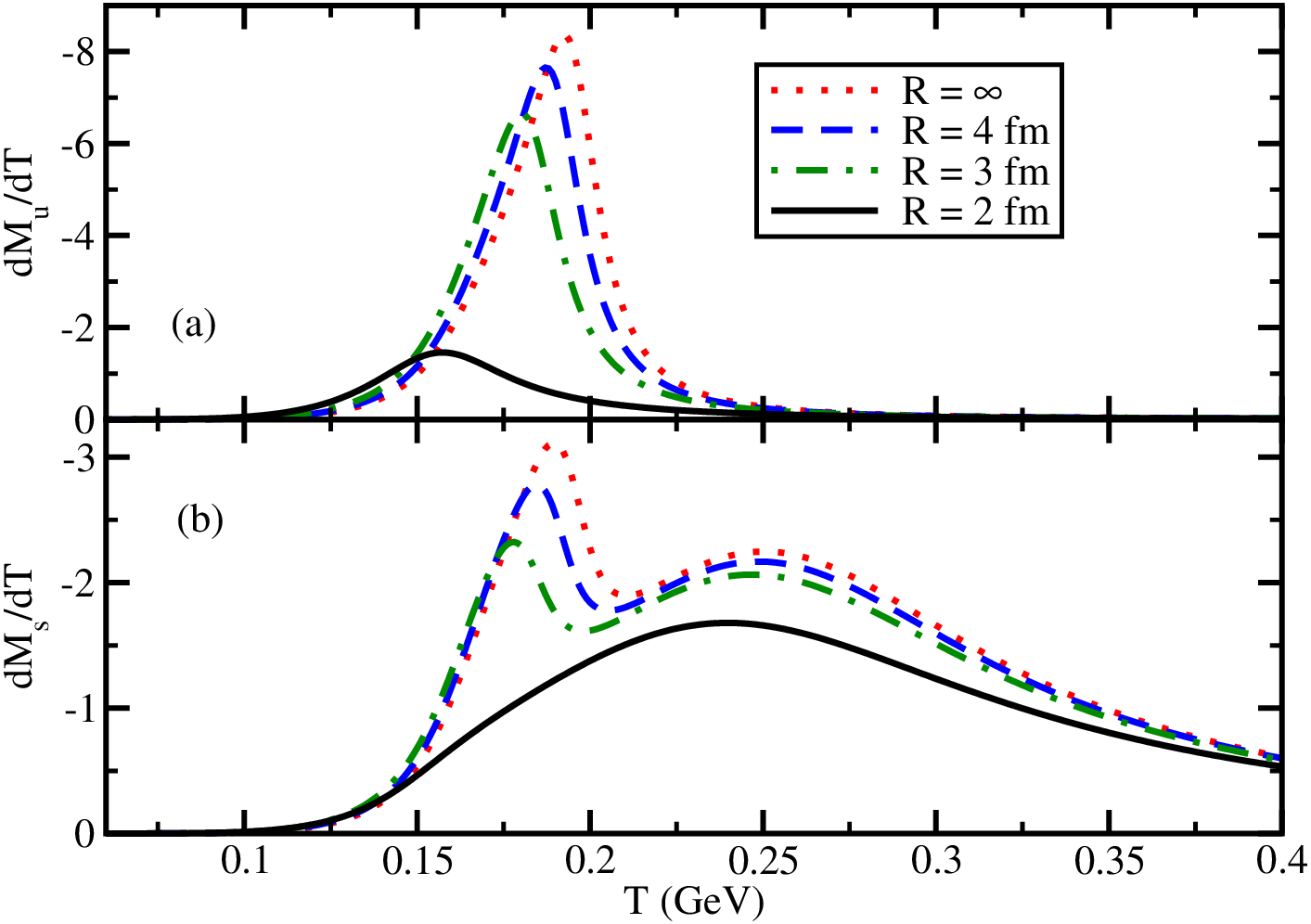}
\caption{(Color online) (a) $\frac{dM_u}{dT}$ and (b) $\frac{dM_s}{dT}$ for for 
$R=\infty$ (red dotted line), $4$ fm (blue dashed line),
$3$ fm (green dash-dotted line) and $2$ fm (black solid line).} 
\label{dm_dT_T}
\end{center}
\end{figure}

To understand these facts, we have to 
focus on the conformal breaking term, given in Eq.~(\ref{conf}), where $dM_{u,s}/dT$ 
is one of the main controlling parameters, which is shown in
Fig.~(\ref{dm_dT_T}). We observe that the peak position of $dM_{u,d}/dT$ in 
Fig.~\ref{dm_dT_T}(a), which represents the transition temperature ($T_c$) of chiral
phase, shifts towards lower temperature as $R$ decreases. The peak strength of $dM_{u,d}/dT$
also decreases when $R$ decreases. If we focus on $\zeta(T)$ for light ($u$ and $d$) quark 
matter only, then it follows exactly same pattern of $dM_{u,d}/dT$ i.e. peak strength
of $\zeta(T)$ reduces and its peak position shifts towards lower $T$ as $R$ decreases.
The complex two peak structure comes into picture when we add $s$ quark contribution,
which participates partially in chiral phase transition. Apart from the expected peak
near $T_c$, $dM_s/dT$ exhibits an additional peak at higher temperature and therefore,
we get two peak structure in $\zeta(T)$ for $2+1$ flavor quark matter. The first peak
in $dM_s/dT$ is little sharper than the second one. Both are reduced when $R$ decreases but
at $R=2$ fm, the first peak almost vanishes. So, only second peak survives in $dM_s/dT$
at $R=2$ fm and its magnitude is interestingly comparable to the corresponding magnitude
of $dM_{u,d}/dT$. As a net effect, we get one broadened (not sharp) peak at some intermediate
location between the peak positions of $dM_{u,d}/dT$ and $dM_s/dT$ for $R=2$ fm.
\begin{figure} 
\begin{center}
\includegraphics[width=0.5 \textwidth]{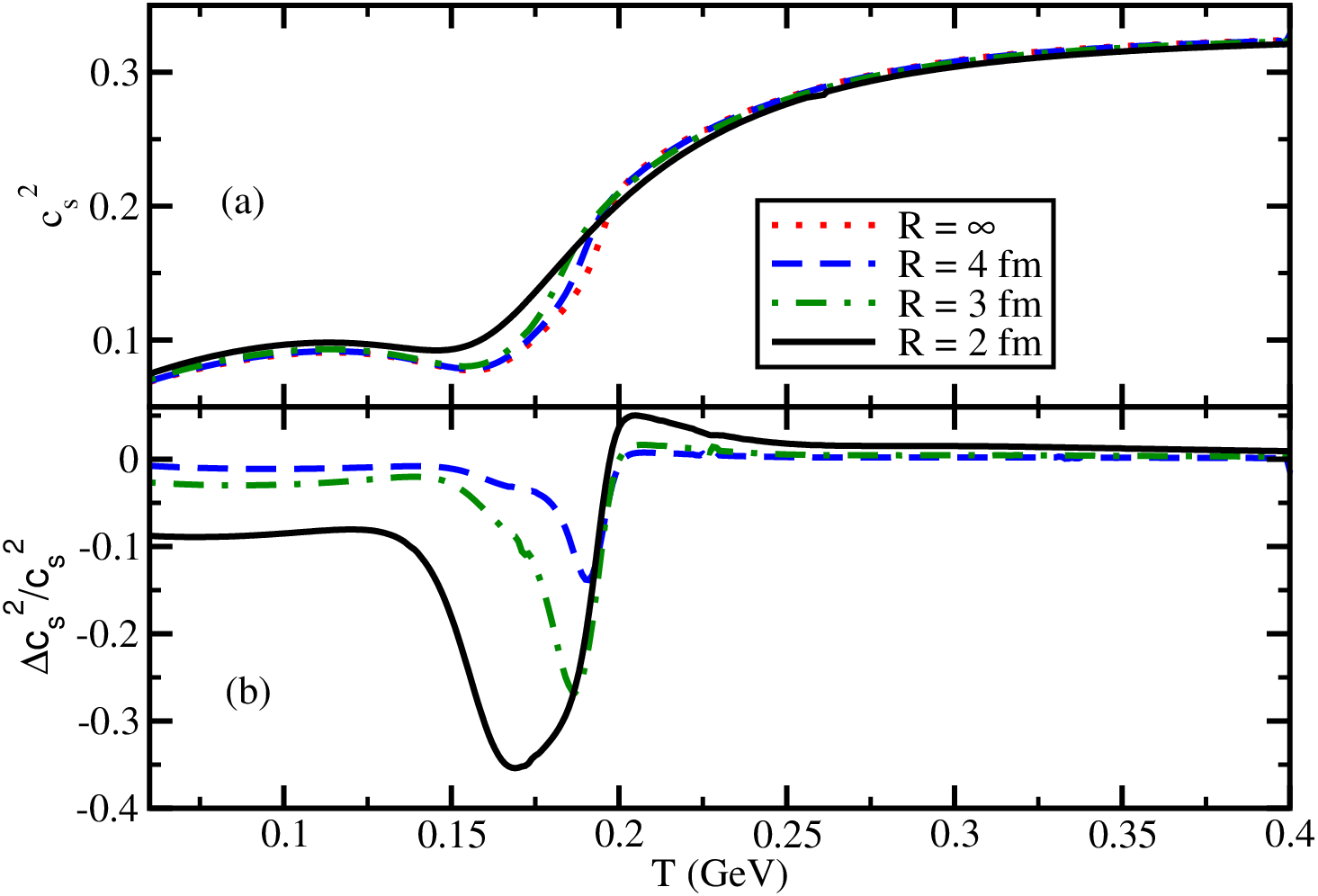}
\caption{(Color online) Same as Fig.~(\ref{zeta_T}) for square of speed of sound $c_s^2$.} 
\label{cs2_T}
\end{center}
\end{figure}
\begin{figure} 
\begin{center}
\includegraphics[width=0.5 \textwidth]{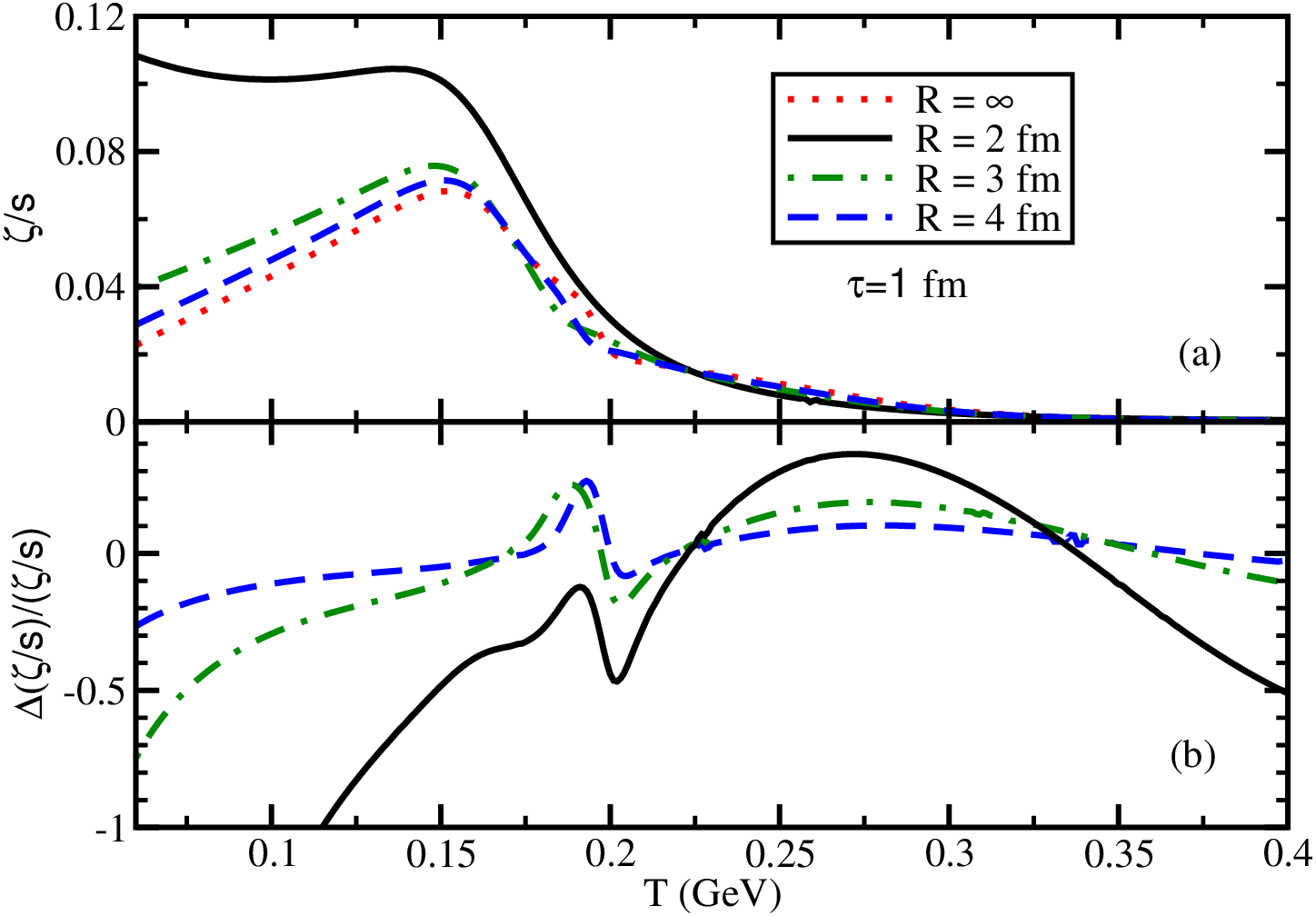}
\caption{(Color online) Same as Fig.~(\ref{zeta_T}) for bulk viscosity to entropy density ratio $\zeta/s$.} 
\label{z_s_T}
\end{center}
\end{figure}
\begin{figure} 
\begin{center}
\includegraphics[width=0.5 \textwidth]{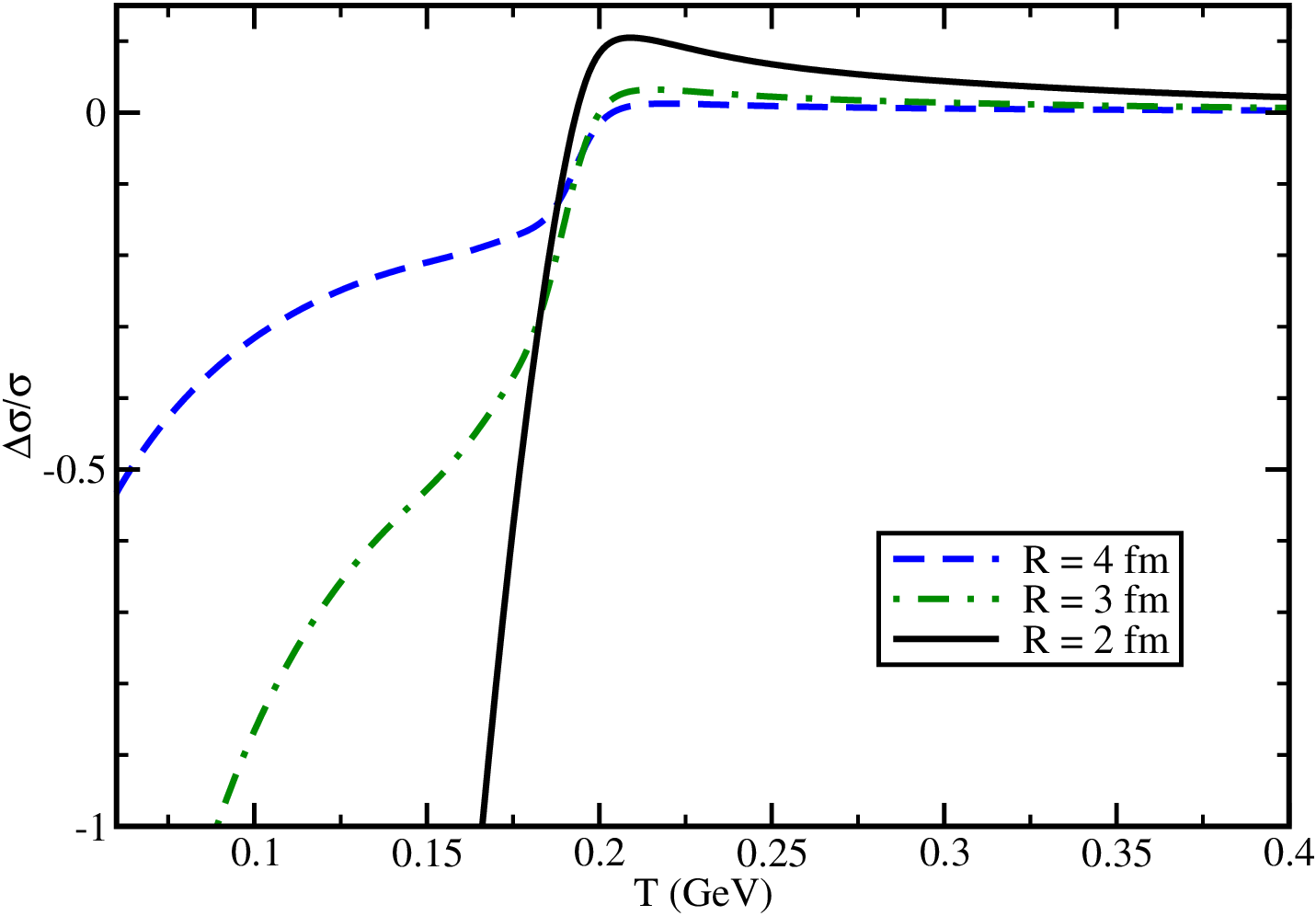}
\caption{(Color online) Difference between finite and infinite matter results for electrical
conductivity $\sigma$ for different values of $R$.} 
\label{cond_T}
\end{center}
\end{figure}

Apart from the term $dM_{u,d,s}/dT$, another component of conformal breaking is $c_s^2$, which
is shown in Fig.~\ref{cs2_T}(a). As $c_s^2=s/\{T~ds/dT\}$, for vanishing quark 
chemical potential the $R$ dependence of $c_s^2$ is quite similar to the $R$ dependence 
of $s$, shown in Fig.~\ref{s_T}(a).
The changes in $c_s^2$ are shown more precisely
in Fig.~\ref{cs2_T} by plotting $\Delta c_s^2/c_s^2$ vs. $T$ for different values of $R$.
So, from $R$ dependences of $dM_{u,d,s}/dT$ and $c_s^2$, given in Figs.~(\ref{dm_dT_T}) and
(\ref{cs2_T}), we can grossly understand the qualitative nature of $\zeta(T,R)$, given
in Fig.~\ref{zeta_T}(b). The changes of $\zeta$ are explored in Fig.~\ref{zeta_T}(b),
where we notice that $\Delta \zeta/\zeta$ remains non-zero at high temperature zone,
unlike $\Delta \eta/\eta$ or others. This is because of strange quark contribution.
Unlike $dM_{u,d}/dT$ curves in Fig.~\ref{dm_dT_T}(a), $dM_{s}/dT$ curves for different 
values of $R$, shown in Fig.~\ref{dm_dT_T}(b), do not merge at high temperature zone 
($T=0.170-0.400$ GeV). If we restrict our outcome to $u$ and $d$ quarks, then
we will get a vanishing $\Delta \zeta/\zeta$ at that high temperature region.

Now, normalizing the bulk viscosity by the entropy density, we have plotted
$\zeta/s$ and its change $\Delta (\zeta/s)/(\zeta/s)$ for different values
of $R$ in Figs.~\ref{z_s_T}(a) and (b) respectively. Though $\zeta$ at
low temperature limit almost tends to zero but $\zeta/s$ at that limit becomes
finite because of their comparable magnitudes. Interestingly, the second (mild)
peak of $\zeta$ almost disappears in $\zeta/s$ because $s(T)$ at high $T$ domain
is strongly dominant and increases rapidly with respect to $\zeta(T)$. Comparing
Figs.~\ref{zeta_T}(b) and \ref{z_s_T}(b), we notice that the changes
of $\zeta$ and $\zeta/s$ due to finite size are approximately similar. 

Next, let us come to the electrical conductivity $\sigma$ of quark matter, whose
expression is given in Eq.~(\ref{sigma_G}). Using this expression, we have generated
$\sigma(T)$ with and without finite size effect and then following our earlier technique,
we define $\Delta\sigma/\sigma$, which is plotted in Fig.~(\ref{cond_T}) for different
values of $R$. As the expressions of $\eta$ and $\sigma$ are quite similar, therefore,
one can see a similarity between the results of Fig.~\ref{eta_T}(b) 
and Fig.~(\ref{cond_T}). However, one should notice that the integrand of 
$\sigma$ is $\vk^2$ times smaller than that of $\eta$, which is the main reason of 
quantitative difference between $\Delta\sigma/\sigma$ and $\Delta\eta/\eta$. For example,
at $T=0.170$ GeV $R=2$ fm, $\Delta\sigma/\sigma=-80\%$ but $\Delta\eta/\eta=-70\%$.
One more interesting difference between $\Delta\sigma/\sigma$ and $\Delta\eta/\eta$ is that
at high $T$ range, $\sigma(R)$ reduces with respect to $\sigma(R=\infty)$. For example,
at $T=0.200$ GeV $R=2$ fm, $\Delta\sigma/\sigma=+8\%$ but $\Delta\eta/\eta=-4\%$.
This clearly happens because of the absence of a $\vec{k}^2$ factor in $\sigma$ compared to that in
$\eta$.
%
\subsection{Comparison with NJL results}
\begin{figure} 
\begin{center}
\includegraphics[width=0.5 \textwidth]{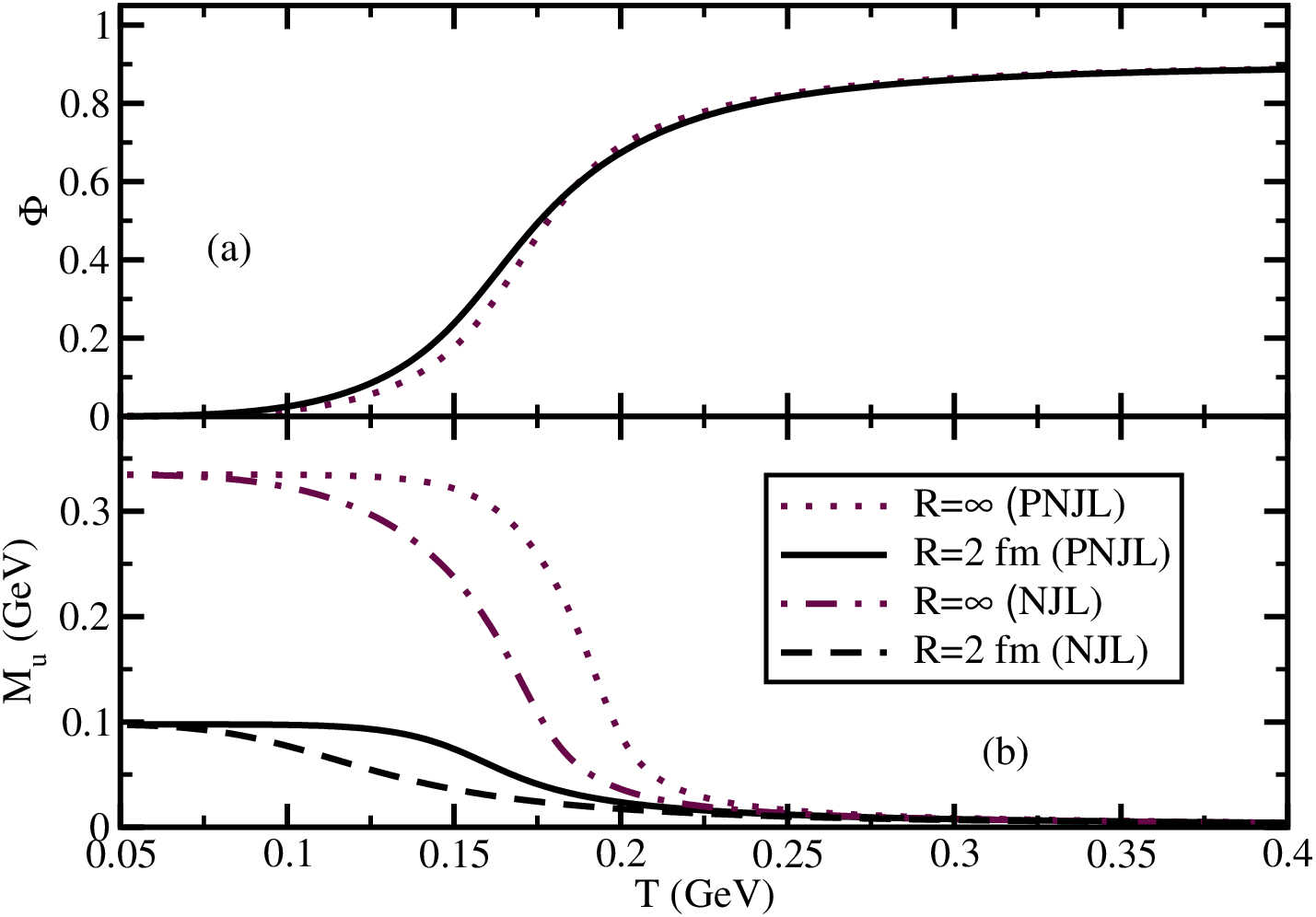}
\caption{(Color online) (a) : $T$ dependence of field $\Phi$ at $R=\infty$ (dotted line) and $R=2$ fm (solid line).
(b) : Quark mass $M_u(T)$ at $R=\infty$ and $R=2$ fm from PNJL model (dotted and solid lines)
and NJL model (dash-dotted and dashed lines).
} 
\label{fiT}
\end{center}
\end{figure}

\begin{figure} 
\begin{center}
\includegraphics[width=0.5 \textwidth]{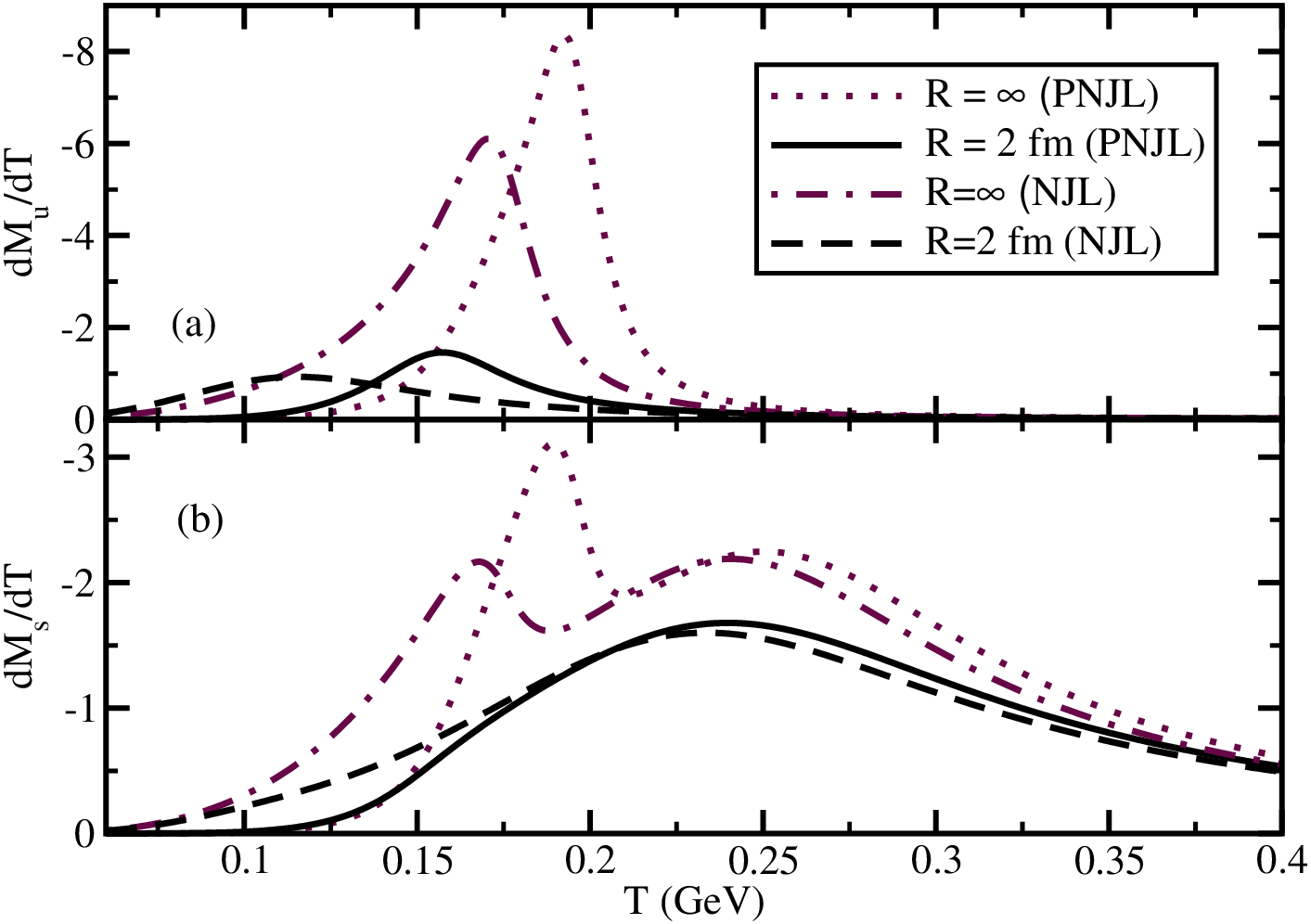}
\caption{(Color online) Same as Fig.~\ref{fiT}(b), the $T$ dependence of $dM_u/dT$ (a) and $dM_s/dT$ (b)
are shown.} 
\label{dm_dT_T2}
\end{center}
\end{figure}

Without Polyakov loop extension, we have also generated the results
for NJL models, where the straight forward Fermi-Dirac distribution function
will describe the statistical probability of quarks in medium. So, the finite size 
effect will enter here through the modification of quark condensates or quark masses
only. Whereas in PNJL model, Polyakov loop field $\Phi$ will also have finite 
size effect along with the condensates. Unlike condensate, the Polyakov
loop in present formalism don't carry any momentum integration so direct implementation
of finite size effect by introducing the lower momentum cutoff is not possible for this case.
However, one can do it by following some different potential, which carry momentum 
integration~\cite{pol1,pol2}. In present framework, Polyakov loop field faces an indirect 
finite size effect.
In Fig.\ref{fiT}(a), we see the field variable
$\Phi$ face a mild change when we jump from $R=\infty$ to $2$ fm, while a drastic change
is noticed in quark mass $M_u$ for the same transition in $R$. Fig.~\ref{fiT}(b)
shows the $T$ dependent quark mass $M_u$ at $R=\infty$ and $R=2$ fm from PNJL model 
(dotted and solid lines) and NJL model (dash-dotted and dashed lines).
We notice that condensate starts to melt down at lower temperature in NJL model than
with respect to PNJL model but the strengths of their condensate at $T=0$ is
exactly same. This observation is true for both infinite and finite matter but
when one transits from infinite to finite matter picture for any model (NJL or PNJL), its 
vacuum condensate strength is abruptly reduced. When we see the curves of Fig.~\ref{dm_dT_T2}(a),
which are basically the temperature derivatives of curves in Fig.~\ref{fiT}(b),
we can visualize the phase transition more clearly. We know that peak position
of $dM_u/dT$ for NJL model gives us the chiral transition temperature while, PNJL model
contain collectively both chiral and deconfinement transition. In our present
model, deconfinement transition takes place at higher temperature than the chiral
transition temperature, therefore, the peak position of $dM_u/dT$ for PNJL model shifts towards
a higher temperature than the same for NJL model. When one transits from infinite to 
finite matter case for any model, the peak position shifts towards lower temperature.
It means that transition temperature decreases by reducing the size of the medium,
which is also noticed in earlier Refs.~\cite{Bhattacharyya:2012rp,Bhattacharyya:2015pra}.
Next, Fig.~\ref{dm_dT_T2}(b) reveals similar kind of peak shifting for strange quark but
it contain an additional hump structure in higher temperature, which is well discussed
in earlier section.
These peak shifting will make direct impact on bulk viscosity. 

\begin{figure} 
\begin{center}
\includegraphics[width=0.5 \textwidth]{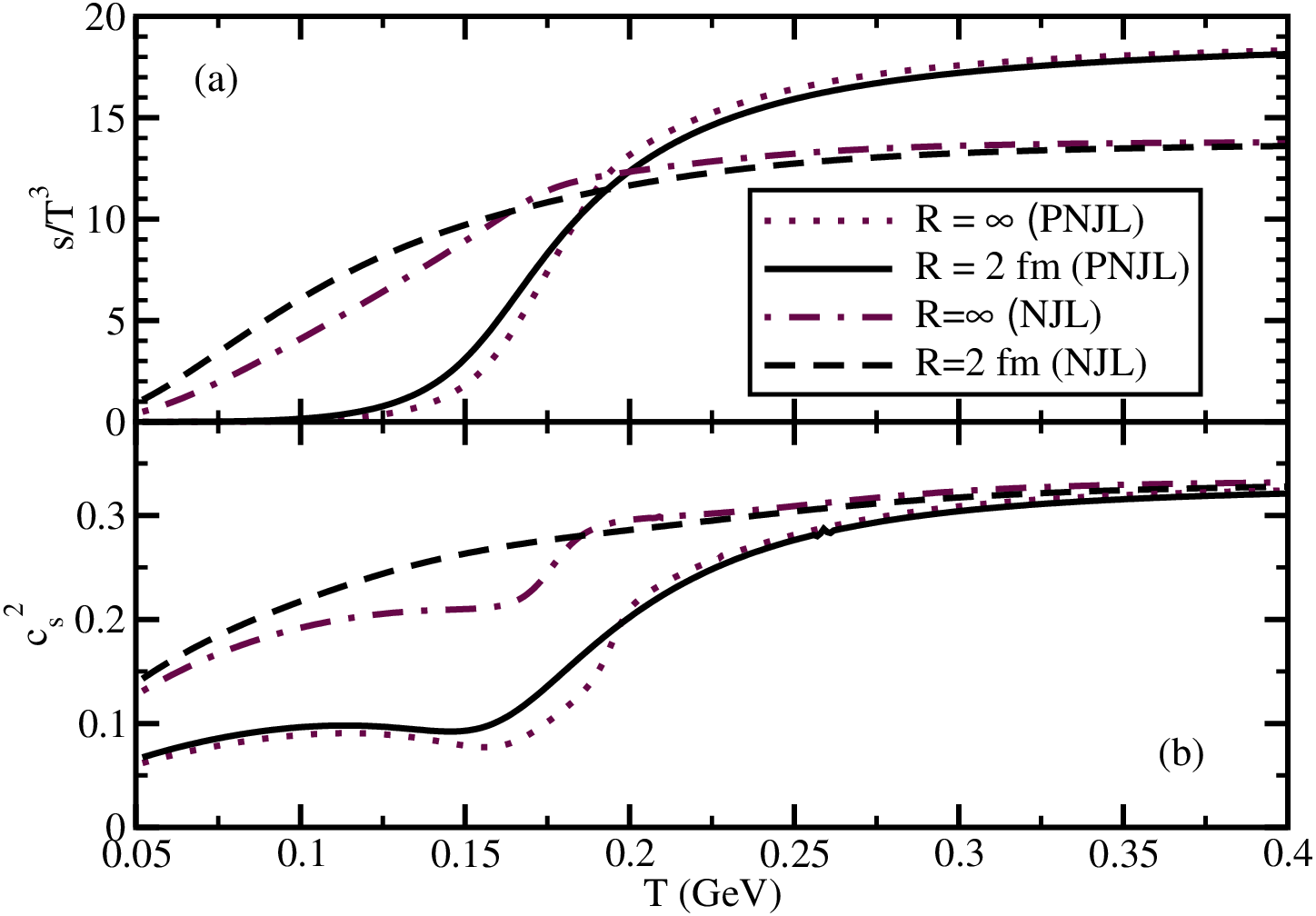}
\caption{(Color online) Following the same pattern of earlier figure, the comparison between
NJL and PNJL model results for entropy density $s(T)$ and speed of sound $c_s(T)$.} 
\label{s_T2}
\end{center}
\end{figure}

\begin{figure} 
\begin{center}
\includegraphics[width=0.5 \textwidth]{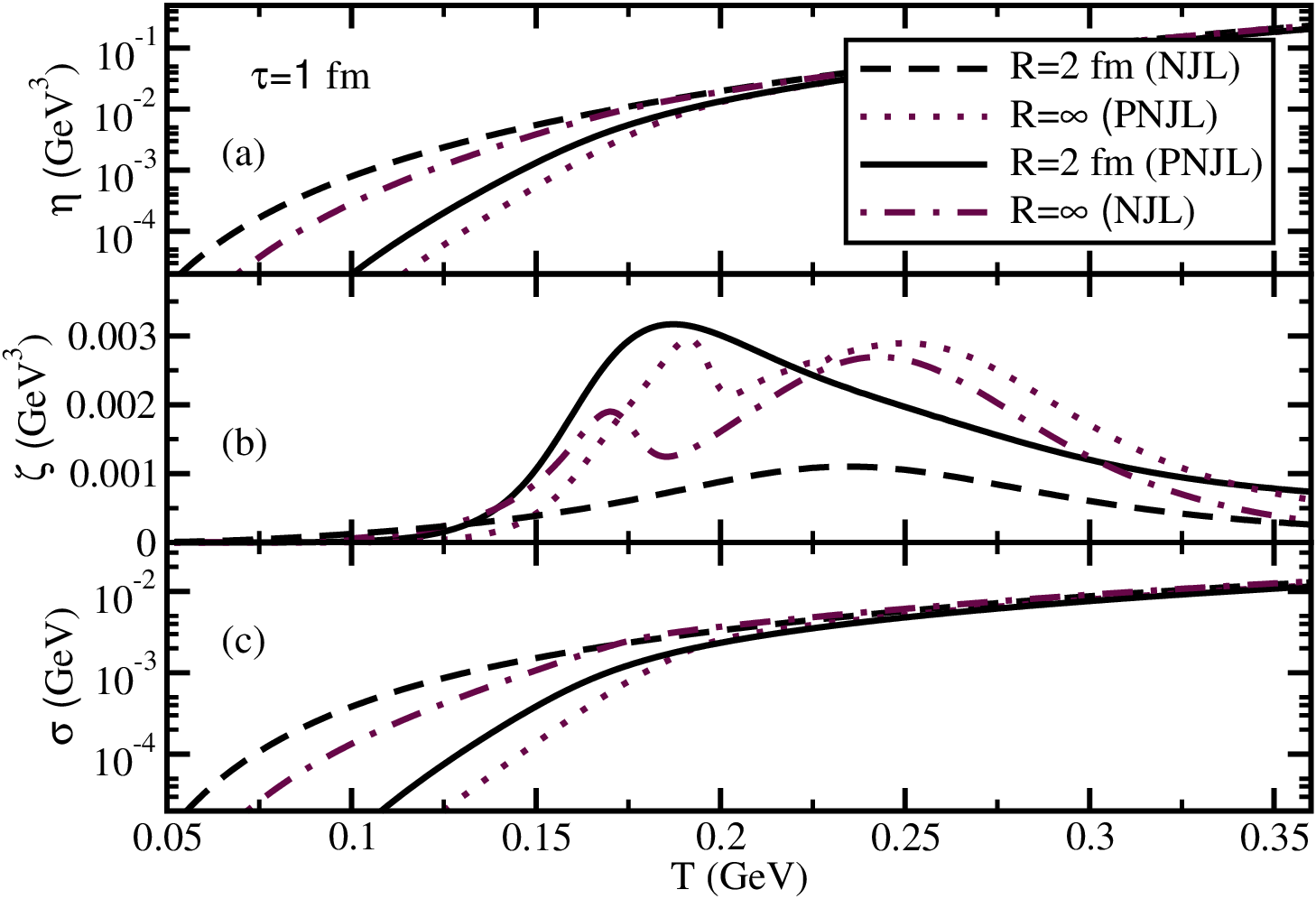}
\caption{(Color online) Following the same pattern of earlier figure, the comparison between
NJL and PNJL model results for shear viscosity $\eta$, bulk viscosity $\zeta$\
and electrical conductivity $\sigma$.} 
\label{NJL_tr}
\end{center}
\end{figure}

Following similar patterns of Figs.~\ref{fiT} and \ref{dm_dT_T2}, the
entropy density $s$ and speed of sound $c_s$ are shown in Figs.\ref{s_T2}(a)
and (b) respectively. In low temperature range, both quantities are enhanced
when one goes from PNJL to NJL model as well as from infinite to finite matter case.
At high temperature, an opposite behavior is observed.
Finally, we come to the transport coefficients - $\eta$, $\zeta$
and $\sigma$ in NJL and PNJL model for $R=\infty$ and $2$ fm, which
are shown in Figs.~\ref{NJL_tr}(a), (b) and (c) respectively.
We can grossly conclude that both $\eta$ and $\sigma$ are enhanced
during the transition from PNJL to NJL model as well as from infinite
to finite matter case. Although, all of the curves merge at high
temperature region. This is expected because the thermal distribution
functions of NJL and PNJL model become same at high temperature range.
Again, the thermodynamic probability of quarks at lower momentum is
negligible for this high temperature domain, so impact of lower momentum
cutoff for finite size consideration of medium, will also be negligible.
Therefore the curves for $R=\infty$ and $R=2$ fm both merge at high
temperature domain.
For bulk viscosity, $dM_u/dT$ and $dM_s/dT$ of 
Figs.~\ref{dm_dT_T2}(a) and (b) will collectively build its profile,
which is little complex in structure. In earlier section, for PNJL model, 
we have already analyzed how two peak structure of $\zeta$ is converted 
to one broadened peak structure. Same event is happening for NJL model
also but its positions of peaks in $T$ axis are only different. 

\subsection{Phenomenological significance}
Now let us see the connection or application of our studies in heavy ion phenomenology.
The expanding medium, created in heavy ion collision experiments,
can be well described by dissipative hydrodynamic simulations, where
the transport coefficients like shear and bulk viscosities are implemented as input
parameters. When the medium expands, its volume increases and temperature
decreases with time. At certain temperature, called freeze-out temperature, 
the medium loses its many body identity. 
In experiment, only this freeze-out size of the medium can be measured.
However, before the freeze-out point, the size of the expanding medium
can be smaller than its freeze-out size. 
Our present investigation reveals that the values of $\eta/s$ and $\zeta/s$ 
can be changed for different system sizes, which are less than 6-7 fm (approx). 
So one should consider size dependent (along with temperature dependent) 
$\eta/s$ and $\zeta/s$ during the complete evolution. 

In most central collision,
the freeze-out size is quite large ($\sim 7-8$ fm) but in 
non-central collision, it can be smaller.
So, at different centrality, one can expect different values
of transport coefficients.
In Ref.~\cite{Roy:2012jb}, the centrality dependence of
invariant yield and elliptic flow of charged hadrons as a
function of transverse momentum has been investigated. 
They have matched the experimental data of PHENIX Collaboration~\cite{PHENIX_pt,PHENIX_v2} 
by taking different guess values of $\eta/s$ in the hydrodynamical simulation
and they found the experimental data
prefers higher values of $\eta/s$ as we go from central to peripheral
collisions. The same indication is found in our present work.
When one goes from central to peripheral collisions, which means from
higher to lower system sizes, our estimated values of $\eta/s$ are enhanced.
From microscopic direction, our understanding is that the quantum effect due to 
finite size of the system is responsible for enhancing the values of $\eta/s$.

\section{Summary and Perspectives} 
\label{sec:summ}

As a first attempt to investigate the qualitative changes brought about by
finite size effect on transport coefficients
of quark matter, we have adopted here a simple idea of taking non-zero lower momentum
cut-off under the framework of PNJL model.
The temperature dependences of constituent quark masses, obtained from the gap
equation, have been modified and they get diminished as size of the system
decreases. When these size dependent quark masses are plugged in the integrands of
different transport coefficients, then some enhancements in their values are found.
Whereas, the expressions of transport coefficients contain momentum integrations as well, whose
non-zero lower limit contributes additionally to the effects of medium size which
basically decreases as we decrease the size. So, there will be a competition between
these two sources of finite size, which will determine the net change in the values
of transport coefficients. In low temperature range, shear viscosity and electrical 
conductivity increase as system size is reduced. The size effect disappears at high
temperature range, as chiral symmetry gets restored there for any system size.
The bulk viscosity, which basically measures the 
scale violation of the medium, has a non-trivial link with the system size. 
The rate
of change of constituent quark mass with respect to temperature and speed of sound
are two quantities responsible for that.

We have also analyzed the same studies for NJL model case just to understand the transition
between NJL and PNJL models. We have noticed that the values of transport coefficients are
grossly enhanced when we transit from PNJL to NJL model as well as from infinite matter
to finite matter. This enhancement mostly occurs in low temperature domain and almost
vanishes at high temperature domain.  

In phenomenological direction, our microscopic calculations say that
$\eta/s$ of the medium increases when one goes from
central to peripheral collisions. Similar conclusion is also found from the macroscopic direction,
where dissipative hydrodynamical simulation describes the expanding medium by taking
$\eta/s$ as an input parameter.

In order to see the qualitative changes in transport coefficients for
finite system size consideration, we have taken constant value of relaxation
time in this present work. However, involved calculations of relaxation time
at finite temperature as well as system sizes incorporating different
interaction channels might lead us to more realistic scenario. We intend
to address the issue in our future project.

{\bf Acknowledgment:} 
SG is partially supported from  University Grant Commission (UGC) 
Dr. D. S. Kothari Post Doctoral Fellowship (India) under
grant No. F.4-2/2006 (BSR)/PH/15-16/0060. KS acknowledges the partial 
financial supports by DST-SERB Ramanujan fellowship of Tamal Mukherjee 
under project no. SB/S2/RJN-29/2013 and DST-SERB NPDF under Fellowship ref. no. PDF/2017/002399. 
SU and SM thank DST and CSIR for financial support. SG acknowledges
WHEPP-2017 for getting some fruitful discussions.

\end{document}